%% file: main.tex
\documentclass[sigconf,natbib=true,anonymous=false,dvipsnames, svgnames]{acmart}
\input{preamble.tex}

\input{tikz/figures.tex}

\pdfoutput=1
\begin{document}

\title{A Static Pruning Study on Sparse Neural Retrievers}


\author{Carlos Lassance}
\email{carlos.lassance@naverlabs.com}
\orcid{0000-0002-7754-6656}
\affiliation{%
  \institution{Naver Labs Europe}
  \city{Meylan}
  \country{France}
}

\author{Hervé Déjean}
\email{herve.dejean@naverlabs.com}
\orcid{0000-0002-9837-5358}
\affiliation{%
  \institution{Naver Labs Europe}
  \streetaddress{6 chemin de Maupertuis}
  \city{Meylan}
    \country{France}
}

\author{Simon Lupart}
\email{simon.lupart@naverlabs.com}
\orcid{0009-0008-2383-4557}
\affiliation{%
  \institution{Naver Labs Europe}
  \city{Meylan}
  \country{France}
}

\author{Stéphane Clinchant}
\email{stephane.clinchant@naverlabs.com}
\orcid{0000-0003-2367-8837}
\affiliation{%
  \institution{Naver Labs Europe}
  \streetaddress{6 chemin de Maupertuis}
  \city{Meylan}
    \country{France}
}

\author{Nicola Tonellotto}
\email{nicola.tonellotto@unipi.it}
\orcid{XXX}
\affiliation{%
  \institution{University of Pisa}
  \city{Pisa}
  \country{Italy}
}



\begin{abstract}

Sparse neural retrievers, such as DeepImpact, uniCOIL and SPLADE, have been introduced recently as an efficient and effective way to perform retrieval with inverted indexes. They aim to learn term importance and, in some cases, document expansions, to provide a more effective document ranking compared to traditional bag-of-words retrieval models such as BM25. However, these sparse neural retrievers have been shown to increase the computational costs and latency of query processing compared to their classical counterparts. To mitigate this, we apply a well-known family of techniques for boosting the efficiency of query processing over inverted indexes: static pruning. We experiment with three static pruning strategies, namely document-centric, term-centric and agnostic pruning, and we assess, over diverse datasets, that these techniques still work with sparse neural retrievers. In particular, static pruning achieves $2\times$ speedup with negligible effectiveness loss ($\leq 2\%$ drop) and, depending on the use case, even $4\times$ speedup with minimal impact on the effectiveness ($\leq 8\%$ drop). Moreover, we show that neural rerankers are robust to candidates from statically pruned indexes.

\end{abstract}



%
\maketitle

\section{Introduction}

Pre-trained language models such as BERT~\cite{devlin2018bert} have been shown to significantly improve the effectiveness of Information Retrieval (IR) systems over traditional ranking models. Dense retrieval is one of the most promising applications of pre-trained language models in IR. In dense retrieval, a pre-trained language model computes dense representations of queries and documents, that must be stored and processed using approximate nearest neighbor algorithms~\cite{JDH17}, preferably on GPUs or TPUs~\cite{tpu}.
However, the ranking of documents in dense retrieval incurs significant computational costs, even if this can be reduced through approaches such as ColBERT~\cite{colbert,eaat}, or by re-ranking only a small set of candidates~\cite{nogueira2019bertranker}. Nevertheless, query processing in dense retrieval is much more expensive than processing documents stored in an inverted index with simpler ranking functions and optimized processing algorithms~\cite{fntir2018}.

Sparse neural retrievers~\cite{unicoil,deepimpactv1,splade} aim to merge these two approaches, using a pre-trained language model to compute very compact document representations, up to a single value, to be stored in an inverted index. In doing so, these retrievers try to reach the effectiveness of the complex pre-trained language models, while retaining the efficiency of simple bag-of-words retrieval models and query processing strategies.
Most sparse neural retrievers integrate two main learning goals: they learn which terms in a document should be indexed (document content learning), and how to represent these terms in a corresponding index (term impact learning). In document content learning, irrelevant terms are removed and relevant terms are added to the documents. During term impact learning, each term in each document is associated with an impact score that will be stored in a posting of the term's posting list. 

Despite the effectiveness and efficiency of sparse neural retrievers w.r.t. more complex neural ranking models, recent works have shown that, while effective retrieval is possible with learned sparse approaches, they are often still much slower than their traditional counterparts~\cite{deepimpactv1}. 
This efficiency loss in neural sparse retrievers has been addressed by introducing modifications to the layout of the posting lists in an inverted index, such as storing both the learned impact and the BM25 score in a single posting~\cite{faster}, splitting a posting list into two separate lists, depending on the impact values of their postings~\cite{mackenzie-etal-2022-accelerating} \car{or accessing documents via clusters for anytime retrieval~\cite{mackenzie2021anytime,mackenzie2023efficient}}. However, the actual content of the inverted index, in terms of terms and postings, does not change. 
A classical solution to increase the efficiency of query processing over inverted indexes is static pruning~\cite{carmel:soffer:2001}. The static pruning of inverted indexes deals with removing information stored in the inverted index to improve the efficiency of query processing with negligible or minimal negative impact on the effectiveness of the retrieval system. \textit{Agnostic} pruning strategies~\cite{carmel:soffer:2001,ntoulas:2007,blanco:2007,chen:2013} focus on removing the low-scoring terms and/or documents from the inverted index. In \textit{term-centric}~\cite{carmel:soffer:2001,demoura:2005,ntoulas:2007,blanco:2010,altingovde:2012} and \textit{document-centric}~\cite{buttcher:2006,thota:2011,altingovde:2012} pruning, the removal of a posting depends on the distribution of the posting scores w.r.t. a posting list and a document, respectively, in which the posting appears in. 

In this work, we investigate the impact of different static pruning strategies on inverted indexes built with sparse neural retrievers. Indeed, the new document representations learned by these systems alter the actual content of documents, and the statistical properties of the corresponding postings lists. \car{We differ from  related previous work~\cite{mdgc20-sigir} as we focus on models that not only perform term weighting but allow for learning the entire retrieval process, also including document and query expansion and query weighting.}  In particular, we investigate four major research questions:

\begin{enumerate}[nolistsep]
    \item Is static pruning effective on sparse neural retrievers on the dataset they were trained on, e.g., MSMARCO?
    \item Can we extend static pruning of sparse neural retrievers to other datasets, e.g., TripClick, even in zero-shot scenarios?
    \item Does static pruning affect the quality of reranking?
    \item Can pruning be combined with other approaches to improve the trade-off between effectiveness and efficiency?
\end{enumerate}
Our experiments demonstrate that static pruning can improve the efficiency of query processing of sparse neural retrievers by a factor of $2\times$ with a negligible loss in effectiveness ($\leq 2\%$) and up to $4\times$ with an acceptable loss of effectiveness ($\leq 8\%$). Moreover, our experiments show that neural rerankers are robust to candidates from statically pruned indexes and that we are able to further push the efficiency/effectiveness first-stage trade-off by combining static pruning with neural network quantization, better pre-trained language models and better distillation data. 

\input{tikz/compare_msmarco/recall/figure2.tex}

\section{Methodology}

\input{methodology.tex}

\section{Experiments}

\input{experiments.tex}

\section{Conclusion}

\input{conclusion.tex}

\begin{acks}
This work is supported, in part, by the spoke ``FutureHPC \& BigData'' of the ICSC – Centro Nazionale di Ricerca in High-Performance Computing, Big Data and Quantum Computing funded by European Union – NextGenerationEU, and  the FoReLab project (Departments of Excellence)
\end{acks}

\bibliographystyle{ACM-Reference-Format}
\bibliography{sample-base}

\input{appendix}

\end{document}

%% file: preamble.tex
\usepackage{enumitem}[nolistsep]
\usepackage{multirow} 
\usepackage{array}
\usepackage[para]{footmisc}
\usepackage{bbm}
\usepackage{adjustbox}
\usepackage{pgfplots}
\usepackage{graphicx}
\usepackage{caption}
\usepackage{subcaption}
\usepackage{pgfkeys}
\usepackage[normalem]{ulem}

\newenvironment{customlegend}[1][]{%
    \begingroup
    \csname pgfplots@init@cleared@structures\endcsname
    \pgfplotsset{#1}%
}{%
    \csname pgfplots@createlegend\endcsname
    \endgroup
}%
\def\addlegendimage{\csname pgfplots@addlegendimage\endcsname}

\usepackage[show]{chato-notes}

\newcommand{\car}[1]{{#1}} 

\AtBeginDocument{%
  \providecommand\BibTeX{{%
    \normalfont B\kern-0.5em{\scshape i\kern-0.25em b}\kern-0.8em\TeX}}}

\setcopyright{acmcopyright}

\copyrightyear{2023}
\acmYear{2023}
\setcopyright{acmlicensed}\acmConference[SIGIR '23]{Proceedings of the 46th
International ACM SIGIR Conference on Research and Development in
Information Retrieval}{July 23--27, 2023}{Taipei, Taiwan}
\acmBooktitle{Proceedings of the 46th International ACM SIGIR Conference on
Research and Development in Information Retrieval (SIGIR '23), July 23--27,
2023, Taipei, Taiwan}
\acmPrice{15.00}
\acmDOI{10.1145/3539618.3591941}
\acmISBN{978-1-4503-9408-6/23/07}



%% file: tikz/figures.tex
\makeatletter
\@namedef{ver@everyshi.sty}{}
\makeatother

\usepackage{pgfplots, pgfplotstable}
\pgfplotsset{compat=newest}
\usepgfplotslibrary{fillbetween}

\newcommand{\spladelargec}{Blue}
\newcommand{\splademediumc}{PineGreen}
\newcommand{\spladesmallc}{RubineRed}
\newcommand{\deepimpacc}{Salmon}
\newcommand{\unicoilc}{Gray}
\newcommand{\bmvc}{Black}

\pgfmathsetmacro{\stdgrad}{30}

\tikzset{every mark/.append style={solid}}
\pgfplotsset{
	grid=both, width=\linewidth, try min ticks=5,
	legend cell align=left, legend style={fill opacity=0.8},
	ylabel near ticks,
    xlabel near ticks,
    every tick label/.append style={font=\footnotesize},
}

\pgfplotsset{
    largespladeTplot/.style={thick, color=\spladelargec, mark=triangle*, mark size=2pt},
    deepimpacTplot/.style={thick, color=\deepimpacc, mark=triangle*, mark size=2pt},
    unicoilTplot/.style={thick, color=\unicoilc, mark=triangle*, mark size=2pt},
    bmvTplot/.style={thick, color=\bmvc, mark=triangle*, mark size=2pt},
    largespladeDplot/.style={thick, color=\spladelargec, mark=o, mark size=2pt},
    mediumspladeDplot/.style={thick, color=\splademediumc, mark=o, mark size=2pt},
    smallspladeDplot/.style={thick, color=\spladesmallc, mark=o, mark size=2pt},
    deepimpacDplot/.style={thick, color=\deepimpacc, mark=o, mark size=2pt}, 
    bmvDplot/.style={thick, color=\bmvc, mark=o, mark size=2pt},
    unicoilDplot/.style={thick, color=\unicoilc, mark=o, mark size=2pt},
    largespladeVplot/.style={thick, color=\spladelargec, mark=x, mark size=2pt},
    mediumspladeVplot/.style={thick, color=\splademediumc, mark=x, mark size=2pt},
    smallspladeVplot/.style={thick, color=\spladesmallc, mark=x, mark size=2pt},
    deepimpacVplot/.style={thick, color=\deepimpacc, mark=x, mark size=2pt}, 
    unicoilVplot/.style={thick, color=\unicoilc, mark=x, mark size=2pt},
    bmvVplot/.style={thick, color=\bmvc, mark=x, mark size=2pt},
    }

%% file: tikz/compare_msmarco/recall/figure2.tex
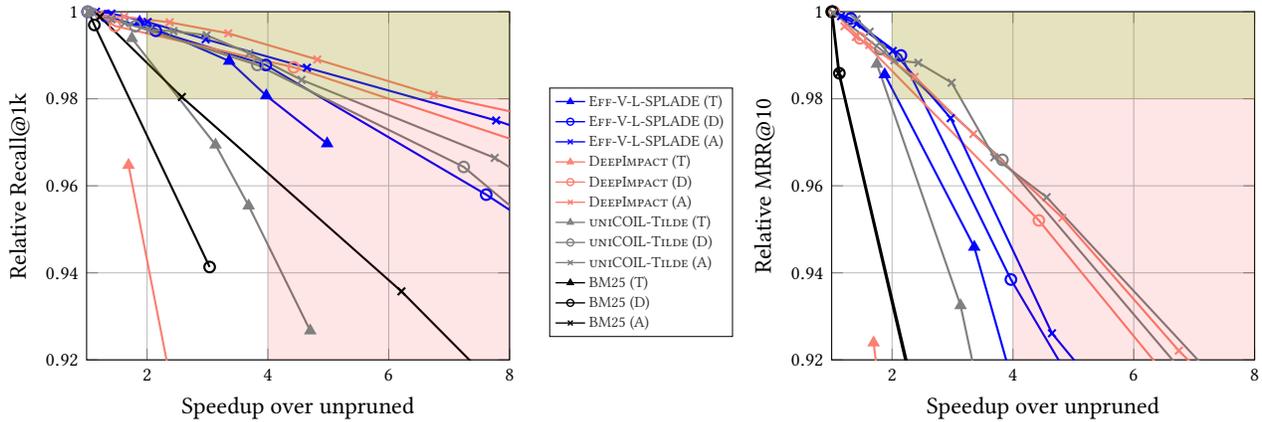
\begin{figure*}[ht]
     \centering
     \begin{subfigure}[hy]{0.85\columnwidth}
     \adjustbox{max width=\textwidth}{%
         \input{tikz/compare_msmarco/recall/pisa.tex}
     }
     \end{subfigure}
     \begin{subfigure}[ht]{0.3\columnwidth}
     \adjustbox{max width=\textwidth}{%
         \input{tikz/compare_msmarco/recall/legend3.tex}
     }
     \end{subfigure}
     \begin{subfigure}[ht]{0.85\columnwidth}
     \adjustbox{max width=\textwidth}{%
         \input{tikz/compare_msmarco/mrr/pisa.tex}
     }
     \end{subfigure}
      
      \caption{Speedup vs relative effectiveness comparison. Rectangles show the two desired conditions (Green $\geq 2\times$ speedup with $\leq 2\%$ effectiveness drop, Red $\geq 4\times$ speedup with $\leq 8\%$ effectiveness drop).}
     \label{fig:relative_effectiveness}
\end{figure*}

%% file: tikz/compare_msmarco/recall/pisa.tex
   \begin{tikzpicture}
       \begin{axis}[
           xlabel=Speedup over unpruned, ylabel= Relative Recall@1k,
           xmin=1.0, xmax=8.0, ymin=0.92, ymax=1.0,
           ]
         \addplot[largespladeTplot] table {tikz/compare_msmarco/recall/splade_T.txt};
         \addplot[unicoilTplot] table {tikz/compare_msmarco/recall/unicoil_T.txt};
         \addplot[deepimpacTplot] table {tikz/compare_msmarco/recall/deep_impact_T.txt};
         \addplot[bmvTplot] table {tikz/compare_msmarco/recall/bm25_T.txt};
         \addplot[largespladeDplot] table {tikz/compare_msmarco/recall/splade_D.txt};
         \addplot[deepimpacDplot] table {tikz/compare_msmarco/recall/deep_impact_D.txt};
         \addplot[unicoilDplot] table {tikz/compare_msmarco/recall/unicoil_D.txt};
         \addplot[bmvDplot] table {tikz/compare_msmarco/recall/bm25_D.txt};
         \addplot[largespladeVplot] table {tikz/compare_msmarco/recall/splade_V.txt};
         \addplot[deepimpacVplot] table {tikz/compare_msmarco/recall/deep_impact_V.txt};
         \addplot[unicoilVplot] table {tikz/compare_msmarco/recall/unicoil_V.txt};
         \addplot[bmvVplot] table {tikz/compare_msmarco/recall/bm25_V.txt};

         \fill[nearly transparent, olive] (2,0.98) rectangle (8,1.0);
         \fill[very nearly transparent, red] (4,0.92) rectangle (8,0.98);
         
       \end{axis}
    \end{tikzpicture}

%% file: tikz/compare_msmarco/recall/legend3.tex
\begin{tikzpicture}
\begin{customlegend}[
legend columns=1,
legend entries={
\textsc{Eff-V-L-SPLADE (T)},
\textsc{Eff-V-L-SPLADE (D)},
\textsc{Eff-V-L-SPLADE (A)},
\textsc{DeepImpact (T)},
\textsc{DeepImpact (D)},
\textsc{DeepImpact (A)},
\textsc{uniCOIL-Tilde (T)},
\textsc{uniCOIL-Tilde (D)},
\textsc{uniCOIL-Tilde (A)},
\textsc{BM25 (T)},
\textsc{BM25 (D)},
\textsc{BM25 (A)}
}]
         \addlegendimage{largespladeTplot};
         \addlegendimage{largespladeDplot};
         \addlegendimage{largespladeVplot};
         \addlegendimage{deepimpacTplot};
         \addlegendimage{deepimpacDplot};
         \addlegendimage{deepimpacVplot};
         \addlegendimage{unicoilTplot};
         \addlegendimage{unicoilDplot};
         \addlegendimage{unicoilVplot};
         \addlegendimage{bmvTplot};
         \addlegendimage{bmvDplot};
         \addlegendimage{bmvVplot};
        \end{customlegend}
\end{tikzpicture}

%% file: tikz/compare_msmarco/mrr/pisa.tex
   \begin{tikzpicture}
       \begin{axis}[
           xlabel=Speedup over unpruned, ylabel= Relative MRR@10,
           xmin=1.0, xmax=8.0, ymin=0.92, ymax=1.0,
           ]
         \addplot[largespladeTplot] table {tikz/compare_msmarco/mrr/splade_T.txt};
         \addplot[unicoilTplot] table {tikz/compare_msmarco/mrr/unicoil_T.txt};
         \addplot[deepimpacTplot] table {tikz/compare_msmarco/mrr/deep_impact_T.txt};
         \addplot[bmvTplot] table {tikz/compare_msmarco/mrr/bm25_T.txt};
         \addplot[largespladeDplot] table {tikz/compare_msmarco/mrr/splade_D.txt};
         \addplot[deepimpacDplot] table {tikz/compare_msmarco/mrr/deep_impact_D.txt};
         \addplot[unicoilDplot] table {tikz/compare_msmarco/mrr/unicoil_D.txt};
         \addplot[bmvDplot] table {tikz/compare_msmarco/mrr/bm25_D.txt};
         \addplot[largespladeVplot] table {tikz/compare_msmarco/mrr/splade_V.txt};
         \addplot[deepimpacVplot] table {tikz/compare_msmarco/mrr/deep_impact_V.txt};
         \addplot[unicoilVplot] table {tikz/compare_msmarco/mrr/unicoil_V.txt};
         \addplot[bmvVplot] table {tikz/compare_msmarco/mrr/bm25_V.txt};
         \fill[nearly transparent, olive] (2,0.98) rectangle (8,1.0);
         \fill[very nearly transparent, red] (4,0.92) rectangle (8,0.98);
       \end{axis}
    \end{tikzpicture}

%% file: methodology.tex
\label{sec:method}
In the following paragraphs, we define the static pruning methods that we use throughout our experiments, \car{and we refer the reader to the supplementary material for a summary of more complex methods that did not improve over the reported ones}. These methods can be divided into term-centric (T), document-centric (D), and agnostic (A) pruning. Each method is tested separately to verify its general behavior and assess the potential advantages of a given approach over the other. 

\paragraph{Remove low scores from posting lists based on score quantile \emph{(T)}} In T we choose a given quantile for each posting list and remove all values that are smaller than that quantile. For example, if we pick 50\% quantile we remove all values equal or smaller than the median value stored on the posting list. In more detail, we test the following quantiles: 50\%, 75\%, 80\%, 85\%. 
The argument for choosing such a method is: if a document-term pair has a score that is on the low end of its respective posting list, there is a small chance for it to be important in score computation, especially if we look into measures with small cutoff values, e.g., MRR@10.

\paragraph{Remove low scores from documents based on a max size \emph{(D)}:} This strategy retains, for each document, the 4, 8, 16, 32, 64 terms with the highest impact values. While in T, pruning is made on the terms posting lists, in D it is made on documents representations before creating posting lists. The argument for the strategy is: the less important terms of the document should be the less useful ones.

\paragraph{Prune low scores \emph{(A)}} Finally, we also verify an approach that is agnostic to both terms and documents, where we simply remove the lowest scores of the inverted index. The idea of this approach is almost the same as for the quantiles-based pruning T, but simplifies the application as it uses the same threshold for all terms. It can also be seen as an extension of L1-regularization-based approaches~\cite{paria2020Minimizing}, \car{ as A prunes values that L1 would push close, but not equal, to 0}. 

%% file: experiments.tex
We perform different experiments to assess if static pruning still works  for the  readily available  sparse neural retrievers, such as 
DeepImpact~\cite{deepimpactv1}, uniCOIL with TILDEv2 expansion~\cite{tildetwo}, and the eff-V-\{S/M/L\} SPLADE family~\cite{lassance2022efficiency}, together with a BM25 baseline. \car{Experimental details are made available in the Appendix and at the codebase\footnote{URL to github}.}
Statistical significance is measured using paired $t$-tests with $p\leq 0.05$ and Bonferroni's correction.

The sparse neural retrievers we evaluate vary in some design choices, i.e.,  i) vocabulary, and ii) expansion. DeepImpact uses a white-space tokenized collection vocabulary with documents being expanded via DocT5~\cite{docT5}, but it does not modify the queries. UniCoil uses the sub-word tokenized BERT vocabulary, with documents expanded using TILDEv2~\cite{tildetwo} and reweighted queries without expansion. Finally, SPLADE also uses the BERT vocabulary but learns both query and document expansions during training using a uniformity regularizer to minimize a \emph{FLOPS} metric~\cite{lassance2022efficiency,paria2020Minimizing}.

\subsection{Static Pruning on MSMARCO}
To verify that the pruning methods proposed in Section~\ref{sec:method} can be extended to sparse neural retrievers, we apply them on the MSMARCO passage dataset~\cite{bajaj2016ms}. 
In this set of experiments, models are both trained and evaluated on MSMARCO, on train and ``dev.small'' query sets, respectively. We also extend our evaluation to the query sets from TREC-DL~\cite{trec19,trec20}, namely the 2019 and 2020 editions. 

Figure~\ref{fig:relative_effectiveness} depicts a comparison of speed-up and relative effectiveness (Recall@1k and MRR@10), i.e., the ratio over the unpruned performance, for all models, under all pruning methods.
We note that for BM25 (black) all pruning strategies are too aggressive, with only a few points appearing before the effectiveness decreased too much. About the different pruning methods, the term-centric pruning (T) works for the models with BERT vocabulary, i.e., Unicoil (grey) and SPLADE (blue), but it is too aggressive for the DeepImpact model (light red), exploiting a larger vocabulary. For the other two pruning methods, i.e., document-centric and agnostic pruning, the results are very similar, with agnostic pruning (cross marker) having a slight advantage, but it needs finer calibration depending on the model as the term impact statistics are different between models. 
Finally, we note that overall the models that do not learn to perform document expansion, but use a predefined expansion, i.e. DeepImpact and Unicoil, generate inverted indexes that look easier to prune. In particular, for SPLADE, stronger pruning entails larger MRR@10 losses. In summary, all neural sparse retrievers we test can get a $2\times$ speedup due to pruning without meaningful loss of effectiveness ($\leq 2\%$ drop in effectiveness) and a $4\times$ speedup is possible with negligible impact ($\leq 8\%$ drop), represented by the green and red rectangles. As we verified that all three methods reported similar trade-offs and that it worked for all models, from now on we shift the focus to document-centric pruning over SPLADE, in order to make the presentation more concise.\footnote{\car{Full-scale experiments can be found in the Appendix}.} 

In Table~\ref{tab:trec}, we look into the TREC-DL (2019 and 2020) query sets~\cite{craswell2020overview,craswell2021overview}. The TREC-DL query sets have fewer queries than the MSMARCO dev set, but have a richer annotation set. Furthermore, the same results still holds: static pruning allows for around $2\times$ speedup (\car{max size of 64}) without meaningful loss of effectiveness ($\leq$ 2\% drop) and around $4\times$ speedup (\car{max size between 16 and 32}) is possible with negligible impact ($\leq$ 8\% drop).

\input{tables/trec.tex}

Finally, we take advantage of the fact that the SPLADE models come in three ``sizes'', namely (S)mall, (M)edium and (L)arge, and look if it is better to generate sparser retrievers by regularization or if pruning by itself is sufficient for achieving a good trade-off between efficiency/effectiveness. The results in Figure~\ref{fig:splades} for Recall@1k show that all sizes are \textit{ Pareto optimal at some point}, in particular for high recall values, which means that L1/FLOPS regularization at training time still plays a role for SPLADE models. The same experiment with the MRR@10 (not shown here) leads to a slightly different conclusion: \emph{pruning the medium-sized SPLADE is overall better}, but it could be explained by the fact that both the medium and large versions have equivalent MRR@10 values, i.e. the MRR@10 is already saturated for the (M)edium version.

\input{tikz/compare_splades/figure.tex}

\subsection{Static Pruning beyond MSMARCO}

In the previous experiments we validated that static pruning is effective for sparse neural retrievers on MSMARCO. However, there are still two questions about its effectiveness: i) does this technique generalize to other datasets? ii) does static pruning of sparse neural retrieval depend on the fact that the training and test sets use the same document collection? In this section, we aim to answer these questions using the TripClick dataset \cite{rekabsaz2021fairnessir}. 

\paragraph{Static pruning generalizes to TripClick:}
The experiments below use the best model described in \cite{pretraining_scratch_arxiv23} for Tripclick, for which we obtained the models from the authors. As shown in the 3rd column in Table~\ref{tab:tripclick}, the pruning effect has even a positive effect on the MRR@10: by pruning documents down to 16 tokens, the effectiveness of the pruned index is similar to the baseline, but with a $4.6\times$ speedup. We observe the same phenomena whatever the query category (high, medium, low frequency), but due to a lack of space, we do not provide these figures. Nevertheless, we note that recall at 1000 is more impacted than MRR at 10: all pruned versions show a statistical significant reduction compared to the baseline.

\input{tables/tripclick.tex}

\paragraph{Static pruning works even on zero-shot evaluation} We also conduct a zero-shot evaluation using the SPLADE model previously finetuned with MSMARCO and evaluated on the TripClick dataset. Results reported in the last three columns of Table~\ref{tab:tripclick} show again similar results to our previous experiments: we obtain $4\times$ speedup with a small impact on effectiveness ($\leq$ 8\% drop on MRR@10).

\subsection{Impact of First Stage Pruning for Reranking}

We now look into what happens if we use the pruned first stage as the input for candidate reranking. The goal is not to reduce the time of the overall system (first stage + reranker), but to look into what would happen if you add a reranker to an already existing pruned index. We use a reimplementation of RankT5~\cite{zhuang2022rankt5} to perform this step, over different pruned versions of eff-V-SPLADE-L and present the results for reranking 1000 candidates in Table~\ref{tab:reranking}. We note that the results are pretty much similar to what we observed before, $2\times$ \emph{first-stage} speedup (size=32) has almost no loss of effectiveness, while for $3\times$ to $4\times$ \emph{first-stage} speedup (size=8), drops in effectiveness are still small enough ($\leq 8\%$), thus we show that the effectiveness of the reranker does not drop when it is added over a pruned index. 

\input{tables/reranking.tex}

\subsection{A More Efficient and Effective SPLADE}

Having shown that static pruning can improve the first stage retrievers, we look into how pruning could be combined with other common tools in order to reduce total latency and improve effectiveness.\car{ In summary}, we start from the eff-V-SPLADE L~\cite{lassance2022efficiency}, which has 0.388 MRR@10 and a total latency of 73 ms (28 retrieval + 45 inference) and aim at greatly reducing latency while keeping a similar MRR@10: i) we use an ensemble of rerankers to generate the distillation scores and change the document encoder to CoCondenser~\cite{gao2021unsupervised}, improving MRR@10 from 0.388 to 0.406, with a slight increase in retrieval latency from 28 to 32 ms; ii) recent improvements from model compilation and quantization~\cite{bai2019} allow us to reduce average inference latency from 45ms to 7ms; iii) document-centric pruning reduces the retrieval latency from 32 to 12 ms ($2.7\times$) for an effectiveness of 0.396 MRR@10 ($\downarrow$ 2.5\%). 

In summary, the new model has better effectiveness than the best model presented in~\cite{lassance2022efficiency}, while having a total latency of 19ms (12ms retrieval + 7ms inference) which is comparable to one of the most efficient models from~\cite{lassance2022efficiency}, which has an effectiveness of 0.376 MRR@10 and a total latency of 13ms (12 ms retrieval + 1ms inference). In other words, we show that pruning can be integrated alongside other improvements to yield state-of-the-art results.

%% file: tables/trec.tex
\begin{table}[ht]
\caption{Experiments on \car{document-centric} pruning of eff-V-L SPLADE on the TREC-DL queries (2019 and 2020). The $\times$ value represents the speedup compared to the baseline. $\dagger$ indicates statistical significance compared to the baseline.}
\label{tab:trec}
\adjustbox{max width=\columnwidth}{%
\begin{tabular}{ccccccc}
\toprule
Max size & \multicolumn{3}{c}{TREC-DL 2019}      & \multicolumn{3}{c}{TREC-DL 2020}   \\
\cmidrule(lr){2-4}\cmidrule(lr){5-7}
 & Latency ($\times$) & nDCG@10 & R@1k & Latency (x) & nDCG@10 & R@1k \\
\midrule
\car{30 522}    & 28.2 (1.0$\times$) & 71.4 & 84.8 & 30.2 (1.0$\times$) & 71.5 & 86.2 \\
\midrule
\multicolumn{7}{c}{Prune by document size (D)} \\
\midrule
64        & 13.2 (2.1$\times$) & 69.3$^\dagger$ & 83.5$^\dagger$  & 13.7 (2.2$\times$) & 70.4 & 84.8$^\dagger$\\
32        & 8.8 (3.2$\times$)  & 67.9$^\dagger$ & 79.0$^\dagger$  & 8.9 (3.4$\times$) & 67.2$^\dagger$ & 81.8$^\dagger$\\
16         & 6.0 (4.7$\times$)  & 64.4$^\dagger$ & 74.6$^\dagger$  & 6.5 (4.6$\times$) & 65.0$^\dagger$ & 79.4$^\dagger$\\
8         & 3.4 (16.6$\times$) & 58.82$^\dagger$ & 67.0$^\dagger$  & 3.5 (8.6$\times$) & 59.5$^\dagger$ & 70.2$^\dagger$\\
4         & 1.7 (16.6$\times$) & 52.4$^\dagger$ & 54.5$^\dagger$  & 1.7 (17.8$\times$) & 48.6$^\dagger$ & 60.7$^\dagger$\\

\bottomrule
\end{tabular}
}
\end{table}



%% file: tikz/compare_splades/figure.tex
\begin{figure}[ht]
     \centering
     \begin{subfigure}[t]{0.75\columnwidth}

         \centering
\adjustbox{max width=\textwidth}{%
         \input{tikz/compare_splades/pisa.tex}
         }
     \end{subfigure}
     \begin{subfigure}[t]{\columnwidth}
         \centering
\adjustbox{max width=\textwidth}{%
         \input{tikz/compare_splades/legend.tex}}
     \end{subfigure}
      \caption{SPLADE models under document-centric pruning. First point on the right of each curve is the unpruned index.}
     \label{fig:splades}
\end{figure}
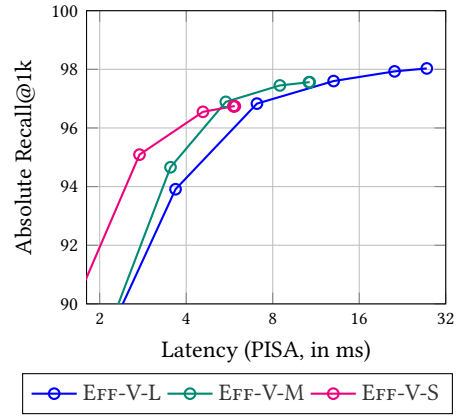

%% file: tikz/compare_splades/pisa.tex
   \begin{tikzpicture}
       \begin{axis}[
           xlabel={Latency (PISA, in ms)},
           ylabel=Absolute Recall@1k,
           ymin=90, ymax=100,
            xmode=log,
            log ticks with fixed point,log basis x={2},
            xmin=1.8,xmax=32,
            xtick={2,4,8,16,32},
                      xticklabels = {2,4,8,16,32}           
           ]
         \addplot[largespladeDplot] table {tikz/compare_splades/splade_large.txt};
         \addplot[mediumspladeDplot] table {tikz/compare_splades/splade_medium.txt};
         \addplot[smallspladeDplot] table {tikz/compare_splades/splade_small.txt};
         
       \end{axis}
    \end{tikzpicture}

%% file: tikz/compare_splades/legend.tex
\begin{tikzpicture}
\begin{customlegend}[
legend columns=3,
legend entries={
\textsc{Eff-V-L},
\textsc{Eff-V-M},
\textsc{Eff-V-S},
                        }]
         \addlegendimage{largespladeDplot};
         \addlegendimage{mediumspladeDplot};
         \addlegendimage{smallspladeDplot};
        \end{customlegend}
\end{tikzpicture}

%% file: tables/tripclick.tex
\begin{table}[ht]
\centering
\caption{
 Experiments on TripClick HEAD queries.$^\dagger$ indicates statistical significance difference compared to the baseline.
}
\adjustbox{max width=\columnwidth}{%
\begin{tabular}{c ccc ccc}
\toprule
\multirow{2}{*}{Max Size}   &    \multicolumn{3}{c}{Finetuning}      & \multicolumn{3}{c}{Zero-Shot}  \\
 \cmidrule(lr){2-4}\cmidrule(lr){5-7}
& Latency ($\times$)      
& MRR@10 
& R@1k   
& Latency ($\times$)      
& MRR@10 
& R@1k   \\ 
\midrule
\car{30 522}     & 5.8 (1.0$\times$) & 54.5 & \textbf{88.8} &  4.2 (1.0$\times$) & \textbf{32.0} & \textbf{83.6} \\
\midrule
\multicolumn{7}{c}{Prune by document size (D)} \\
\midrule


64 &
 3.1 (1.9$\times$)  &
\textbf{56.7}$^\dagger$&
88.3$^\dagger$ &
2.5  (1.7$\times$)  &
31.6\hphantom{$\dagger$}&
82.8$^\dagger$ \\ 
32 & 
1.9 (3.0$\times$) &
55.8\hphantom{$\dagger$}&
86.8$^\dagger$ &
 1.7  (2.5$\times$)  &
31.5\hphantom{$\dagger$}&
80.1$^\dagger$ \\ 
16 &
  1.2 (4.6$\times$)  &
54.7\hphantom{$\dagger$} &
83.4$^\dagger$ &
 1.1  (3.8$\times$) &
29.9$^\dagger$ &
73.4 $^\dagger$ \\ 
8 &
1.0 (5.8$\times$)   &
51.3$^\dagger$&
76.8$^\dagger$ &
1.0  (4.1$\times$)  &
27.7$^\dagger$  &
58.9 $^\dagger$ \\
4 &
 1.0 (5.6$\times$)  &
42.9$^\dagger$ &
66.0$^\dagger$ &
0.8  (5.2$\times$)  &
22.5$^\dagger$ &
41.3$^\dagger$  \\ 
\bottomrule
\end{tabular}
}
\label{tab:tripclick}
\end{table}

%% file: tables/reranking.tex
\begin{table}[ht]
\caption{Reranking eff-V-SPLADE-L with static pruning. Latency considers only the first stage step. $^\dagger$ indicates statistical significant drop compared to the reranked baseline.}
\label{tab:reranking}
\adjustbox{max width=\columnwidth}{%
\begin{tabular}{ccccc}
\toprule
\multirow{2}{*}{Max size}  & \multicolumn{2}{c}{MSMARCO dev}      & \multicolumn{2}{c}{TREC-DL 2019} \\
\cmidrule(lr){2-3}\cmidrule(lr){4-5}
& Latency ($\times$)     & MRR@10 & Latency ($\times$)     & nDCG@10    \\
\midrule
\multicolumn{5}{c}{Baseline without reranking} \\
\midrule
\car{30 522}     & 28.0 (1.0$\times$)           & 38.9\hphantom{$^\dagger$} & 28.2 (1.0$\times$)           & 71.4 \\
\midrule
\multicolumn{5}{c}{Baseline with reranking} \\
\midrule
\car{30 522}     & 28.0 (1.0$\times$)          & 42.8\hphantom{$^\dagger$} & 28.2 (1.0$\times$)           & 77.1 \\
\midrule
\multicolumn{5}{c}{Prune by document size (D)} \\
\midrule
64        & 13.0 (2.2$\times$) & 42.9\hphantom{$^\dagger$} & 14.1 (2.1$\times$) & 77.0 \\
32        & 7.1 (4.0$\times$) & 42.7\hphantom{$^\dagger$} & 10.3 (3.2$\times$) & 75.8 \\
16        & 3.7 (7.6$\times$) & 42.4$^\dagger$ & 8.4 (4.7$\times$) & 76.4 \\
8         & 2.0 (13.9$\times$) & 41.5$^\dagger$ & 5.9 (6.4$\times$) & 75.6 \\
4         & 1.0 (27.0$\times$) & 38.8$^\dagger$ & 3.4 (16.6$\times$) & 74.1 \\
\bottomrule
\end{tabular}
}
\end{table}

%% file: conclusion.tex
In this work, we have studied if basic static pruning is effective given the shift to neural sparse retrievers. We verify experimentally that even the simpler techniques still work and that for most methods we are able to achieve a 2x speedup with practically no loss (less than 2\% loss of MRR@10) and 4x speedup with negligible loss (less than 8\% loss of MRR@10). Furthermore, we have shown that static pruning still works beyond MSMARCO in a zero-shot manner and that efficient pruning does not hurt the reranking performance. 
Finally, by combining static pruning with other techniques (quantization, training improvements), a more effective and efficient SPLADE model can be obtained, thus advancing the state of the art on the efficiency-effectiveness trade-off.

%% file: appendix.tex

\appendix

\section{Negative results with more complex static pruning alternatives}

To start, we would like to note something about our methods, \emph{they are simple}. Indeed, we focus on presenting a concise experimental setup of just 3 pruning techniques and their results for various retrieval methods. One main criticism is: why do we choose only these simple methods, while many others exist, with increasing complexity? In parallel, we actually tested a number of other techniques that did not work as well as the ones we just described, and require deeper investigation. In summary, we also tested: 
\begin{enumerate}
    \item Using \emph{different thresholds} for \emph{expansion and non-expansion terms} (in document-centric and agnostic pruning): Considering that expanded terms are less important compared to the non-expanded ones and thus could be more heavily pruned. We saw almost no difference to a single threshold.
    \item Term-centric pruning with \emph{varying quantiles} depending on the size of the term posting list: In (T) we prune all terms to a given quantile.  Another approach would be to vary the quantiles depending on the posting list length. No improvements were found compared to the simpler (T).
    \item Term-centric pruning to a maximum posting list size: Instead of choosing a quantile of the original posting list, one could use a desired maximum length. The results were worse than with a quantile-based approach.
    \item Reducing the scores during value agnostic pruning: When performing agnostic pruning we change the minimum score from 0 to the threshold score. Another solution would be to reduce all values so that the new threshold equals 0. This approach led to results worse than just keeping the minimum score as the threshold.
\end{enumerate}

\section{Experimental details}

Experiments are performed on a single machine, Intel(R) Xeon(R) Gold 6338 CPU @ 2.00GHz CPU, and evaluated using PISA. Full details for reproducing experiments are made available at the codebase: 
Inference time (i.e., the time to compute query term weights and expansion, if required) is disregarded for most of the experiments, as we only compare within the same retrieval model, thus having the same inference time. Inter-model comparison is only made with relative values. Latency is always reported as the average value across the set of queries in each experiment and measured in milliseconds. 

\input{appendix_tikz/compare_msmarco_index/recall/figure2.tex}

\section{Index size}

In the main paper, we consider solely the retrieval inference time but ignored the index size reduction that is incurred due to pruning. We present the values in Figure~\ref{fig:relative_effectiveness} which is similar to Figure 1 of the main paper, but using relative index size instead of relative latency. We note that the compression effect is less pronounced than the speedup for most retrieval and pruning method combinations, especially for BM25 which is already pretty well compressed. The only exceptions are using SPLADE for retrieval, which we posit is because of the large size of the original index and the use of term-based pruning which makes sense as smaller posting lists are easier to compress.

\section{Full TREC-DL results}

In this Section, we present extensions to Table 1 of the full paper, where we consider all retrievers and all pruning methods. Results are presented in tables~\ref{tab:trec_bm25_full},~\ref{tab:trec_unicoil_full},~\ref{tab:trec_deepimpact_full}, and ~\ref{tab:trec_splade_full}. Overall we see the same results as in the shortened table of the full paper and the MSMARCO figure: we are able to get 2x speedup with $\leq 2\%$ loss of effectiveness and 4x speedup with $\leq 8\%$ loss of effectiveness with pretty much all pruning methods over all learned sparse retrieval models, however, we see that some conclusions could change, as for example, TREC-DL and SPLADE term-based pruning can sometimes be more effective than the other methods and that sometimes evaluation on TREC-DL may be skewed by a few queries, leading to a few results outside of the expected: ``more pruning equals faster and more approximate method''. 

\input{appendix_tables/TREC/bm25}

\input{appendix_tables/TREC/unicoil}

\input{appendix_tables/TREC/deepimpact}

\input{appendix_tables/TREC/splade}

%% file: appendix_tikz/compare_msmarco_index/recall/figure2.tex
\begin{figure*}[ht]
     \centering
     \begin{subfigure}[hy]{0.85\columnwidth}
     \adjustbox{max width=\textwidth}{%
         \input{appendix_tikz/compare_msmarco_index/recall/pisa.tex}
     }
     \end{subfigure}
     \begin{subfigure}[ht]{0.3\columnwidth}
     \adjustbox{max width=\textwidth}{%
         \input{appendix_tikz/compare_msmarco_index/recall/legend3.tex}
     }
     \end{subfigure}
     \begin{subfigure}[ht]{0.85\columnwidth}
     \adjustbox{max width=\textwidth}{%
         \input{appendix_tikz/compare_msmarco_index/mrr/pisa.tex}
     }
     \end{subfigure}
      
      \caption{Index compression vs relative effectiveness comparison. We note that for most methods (safe for SPLADE) the compression effect is less pronounced compared to the speedup.}
     \label{fig:relative_effectiveness}
\end{figure*}
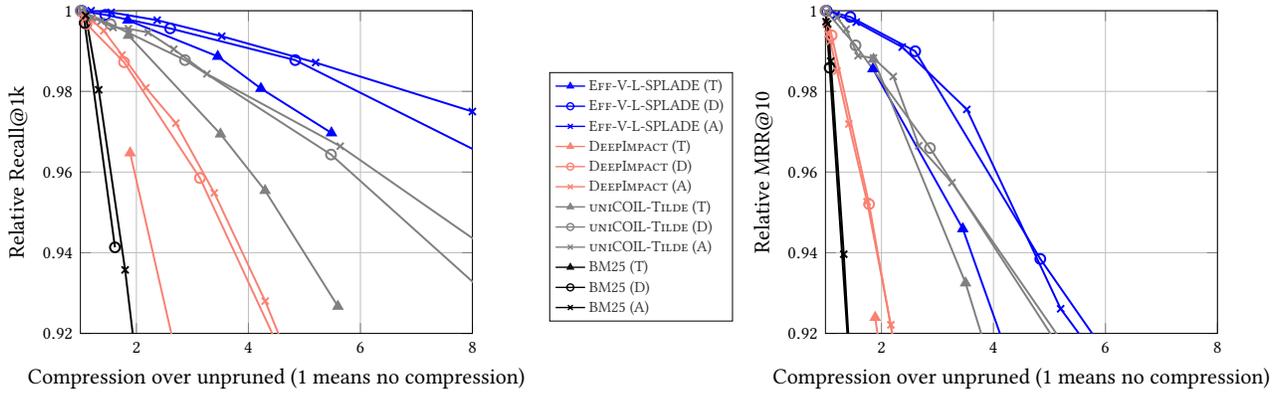

%% file: appendix_tikz/compare_msmarco_index/recall/pisa.tex
   \begin{tikzpicture}
       \begin{axis}[
           xlabel=Compression over unpruned (1 means no compression), ylabel= Relative Recall@1k,
           xmin=1.0, xmax=8.0, ymin=0.92, ymax=1.0,
           ]
         \addplot[largespladeTplot] table {appendix_tikz/compare_msmarco_index/recall/splade_T.txt};
         \addplot[unicoilTplot] table {appendix_tikz/compare_msmarco_index/recall/unicoil_T.txt};
         \addplot[deepimpacTplot] table {appendix_tikz/compare_msmarco_index/recall/deep_impact_T.txt};
         \addplot[bmvTplot] table {appendix_tikz/compare_msmarco_index/recall/bm25_T.txt};
         \addplot[largespladeDplot] table {appendix_tikz/compare_msmarco_index/recall/splade_D.txt};
         \addplot[deepimpacDplot] table {appendix_tikz/compare_msmarco_index/recall/deep_impact_D.txt};
         \addplot[unicoilDplot] table {appendix_tikz/compare_msmarco_index/recall/unicoil_D.txt};
         \addplot[bmvDplot] table {appendix_tikz/compare_msmarco_index/recall/bm25_D.txt};
         \addplot[largespladeVplot] table {appendix_tikz/compare_msmarco_index/recall/splade_V.txt};
         \addplot[deepimpacVplot] table {appendix_tikz/compare_msmarco_index/recall/deep_impact_V.txt};
         \addplot[unicoilVplot] table {appendix_tikz/compare_msmarco_index/recall/unicoil_V.txt};
         \addplot[bmvVplot] table {appendix_tikz/compare_msmarco_index/recall/bm25_V.txt};

         
       \end{axis}
    \end{tikzpicture}

%% file: appendix_tikz/compare_msmarco_index/recall/legend3.tex
\begin{tikzpicture}
\begin{customlegend}[
legend columns=1,
legend entries={
\textsc{Eff-V-L-SPLADE (T)},
\textsc{Eff-V-L-SPLADE (D)},
\textsc{Eff-V-L-SPLADE (A)},
\textsc{DeepImpact (T)},
\textsc{DeepImpact (D)},
\textsc{DeepImpact (A)},
\textsc{uniCOIL-Tilde (T)},
\textsc{uniCOIL-Tilde (D)},
\textsc{uniCOIL-Tilde (A)},
\textsc{BM25 (T)},
\textsc{BM25 (D)},
\textsc{BM25 (A)}
}]
         \addlegendimage{largespladeTplot};
         \addlegendimage{largespladeDplot};
         \addlegendimage{largespladeVplot};
         \addlegendimage{deepimpacTplot};
         \addlegendimage{deepimpacDplot};
         \addlegendimage{deepimpacVplot};
         \addlegendimage{unicoilTplot};
         \addlegendimage{unicoilDplot};
         \addlegendimage{unicoilVplot};
         \addlegendimage{bmvTplot};
         \addlegendimage{bmvDplot};
         \addlegendimage{bmvVplot};
        \end{customlegend}
\end{tikzpicture}

%% file: appendix_tikz/compare_msmarco_index/mrr/pisa.tex
   \begin{tikzpicture}
       \begin{axis}[
           xlabel=Compression over unpruned (1 means no compression), ylabel= Relative MRR@10,
           xmin=1.0, xmax=8.0, ymin=0.92, ymax=1.0,
           ]
         \addplot[largespladeTplot] table {appendix_tikz/compare_msmarco_index/mrr/splade_T.txt};
         \addplot[unicoilTplot] table {appendix_tikz/compare_msmarco_index/mrr/unicoil_T.txt};
         \addplot[deepimpacTplot] table {appendix_tikz/compare_msmarco_index/mrr/deep_impact_T.txt};
         \addplot[bmvTplot] table {appendix_tikz/compare_msmarco_index/mrr/bm25_T.txt};
         \addplot[largespladeDplot] table {appendix_tikz/compare_msmarco_index/mrr/splade_D.txt};
         \addplot[deepimpacDplot] table {appendix_tikz/compare_msmarco_index/mrr/deep_impact_D.txt};
         \addplot[unicoilDplot] table {appendix_tikz/compare_msmarco_index/mrr/unicoil_D.txt};
         \addplot[bmvDplot] table {appendix_tikz/compare_msmarco_index/mrr/bm25_D.txt};
         \addplot[largespladeVplot] table {appendix_tikz/compare_msmarco_index/mrr/splade_V.txt};
         \addplot[deepimpacVplot] table {appendix_tikz/compare_msmarco_index/mrr/deep_impact_V.txt};
         \addplot[unicoilVplot] table {appendix_tikz/compare_msmarco_index/mrr/unicoil_V.txt};
         \addplot[bmvVplot] table {appendix_tikz/compare_msmarco_index/mrr/bm25_V.txt};
       \end{axis}
    \end{tikzpicture}

%% file: appendix_tables/TREC/bm25.tex
\begin{table}[ht]
\caption{Experiments with BM25 on the TREC-DL queries (2019 and 2020). The $\times$ value represents the speedup compared to the baseline.}
\label{tab:trec_bm25_full}
\adjustbox{max width=\columnwidth}{%
\begin{tabular}{ccccccc}
\toprule
Pruning parameter & \multicolumn{3}{c}{TREC-DL 2019}      & \multicolumn{3}{c}{TREC-DL 2020}   \\
\cmidrule(lr){2-4}\cmidrule(lr){5-7}
 & Latency ($\times$) & nDCG@10 & R@1k & Latency (x) & nDCG@10 & R@1k \\
\midrule
None    & 7.6 & 49.9 & 75.5 & 8.1 & 48.5 & 81.1 \\
\midrule
\multicolumn{7}{c}{Prune by document size (D)} \\
\midrule
32        & 6.8 (1.1$\times$) & 48.2 & 75.2 & 7.77 (1.5$\times$) & 48.2 & 80.6 \\
16        & 2.8 (2.7$\times$) & 45.0 & 70.5 & 3.53 (3.4$\times$) & 45.7 & 78.2 \\
\midrule
\multicolumn{7}{c}{Prune low posting list scores (T)} \\
\midrule
Prune lower 50\%        & 4.3 (1.8$\times$) & 43.6 & 60.8 & 5.8 (1.4$\times$) & 43.7 & 63.0 \\
Prune lower 75\%        & 2.6 (2.9$\times$) & 39.7 & 46.1 & 3.4 (2.4$\times$) & 39.2 & 48.6 \\
Prune lower 85\%        & 1.9 (4.0$\times$) & 32.9 & 34.7 & 2.4 (3.4$\times$) & 32.6 & 37.6 \\
\midrule
\multicolumn{7}{c}{Prune low scores (A)} \\
\midrule
Prune $< 10.0$       & 7.9 (2.7$\times$) & 49.8 & 76.4 & 7.2 (2.7$\times$) & 48.7 & 81.1 \\
Prune $< 20.0$       & 6.3 (2.7$\times$) & 48.1 & 76.4 & 7.5 (2.7$\times$) & 46.4 & 80.8 \\
Prune $< 30.0$       & 3.1 (2.7$\times$) & 47.2 & 74.6 & 4.3 (2.7$\times$) & 46.8 & 79  \\
Prune $< 40.0$       & 1.3 (2.7$\times$) & 43.5 & 70.9 & 1.8 (2.7$\times$) & 45.5 & 79.0 \\
\bottomrule
\end{tabular}
}
\end{table}

%% file: appendix_tables/TREC/unicoil.tex
\begin{table}[ht]
\caption{Experiments with uniCOIL-TILDEv2 on the TREC-DL queries (2019 and 2020). The $\times$ value represents the speedup compared to the baseline. Statistical significance not computed.}
\label{tab:trec_unicoil_full}
\adjustbox{max width=\columnwidth}{%
\begin{tabular}{ccccccc}
\toprule
Pruning parameter & \multicolumn{3}{c}{TREC-DL 2019}      & \multicolumn{3}{c}{TREC-DL 2020}   \\
\cmidrule(lr){2-4}\cmidrule(lr){5-7}
 & Latency ($\times$) & nDCG@10 & R@1k & Latency (x) & nDCG@10 & R@1k \\
\midrule
None    & 27.8 & 70.8 & 83.2 & 28.2 & 68.5 & 83.9\\
\midrule
\multicolumn{7}{c}{Prune by document size (D)} \\
\midrule
64        & 14.8 (1.9$\times$) & 69.2 & 82.6 & 15.8 (1.8$\times$) & 68.5 & 82.6 \\
32        & 8.2 (3.4$\times$)  & 67.7 & 80.4 & 7.9 (3.6$\times$)  & 65.2 & 80.1 \\
16        & 5.9 (4.7$\times$)  & 65.2 & 74.9 & 6.7 (4.2$\times$)  & 63.6 & 77.0 \\
8         & 3.7 (7.5$\times$)  & 61.1 & 66.0 & 3.8 (7.4$\times$)  & 58.3 & 66.7 \\
4         & 2.2 (12.6$\times$)  & 49.7 & 51.7 & 2.3 (12.3$\times$)  & 42.8 & 53.4 \\
\midrule
\multicolumn{7}{c}{Prune low posting list scores (T)} \\
\midrule
Prune lower 50\%        & 15.2 (1.8$\times$) & 70.1 & 81.6 & 17.4 (1.6$\times$) & 67.9 & 82.2 \\
Prune lower 75\%        & 8.6 (3.2$\times$)  & 68.5 & 75.8 & 9.0 (3.1$\times$) & 63.2 & 77.0 \\
Prune lower 80\%        & 7.8 (3.6$\times$)  & 67.6 & 72.4 & 8.7 (3.2$\times$) & 60.1 & 75.1 \\
Prune lower 85\%        & 7.6 (3.7$\times$)  & 64.7 & 68.6 & 7.7 (3.7$\times$) & 59.9 & 72.7 \\
\midrule
\multicolumn{7}{c}{Prune low scores (A)} \\
\midrule
Prune $< 0.60$       & 13.6 (2.0$\times$) & 69.0 & 82.5 & 14.8 (1.9$\times$) & 68.6 & 83.1\\
Prune $< 0.80$       & 9.4 (3.0$\times$)  & 67.9 & 82.0 & 9.9 (2.8$\times$)  & 81.3 & 81.3\\
Prune $< 1.00$       & 7.5 (3.7$\times$)  & 66.5 & 80.6 & 7.8 (3.6$\times$)  & 64.6 & 79.8\\
Prune $< 1.25$       & 4.9 (5.7$\times$)  & 65.2 & 75.4 & 6.3 (4.5$\times$)  & 63.2 & 76.7\\
\bottomrule
\end{tabular}
}
\end{table}

%% file: appendix_tables/TREC/deepimpact.tex
\begin{table}[ht]
\caption{Experiments with DeepImpact on the TREC-DL queries (2019 and 2020). The $\times$ value represents the speedup compared to the baseline.}
\label{tab:trec_deepimpact_full}
\adjustbox{max width=\columnwidth}{%
\begin{tabular}{ccccccc}
\toprule
Pruning parameter & \multicolumn{3}{c}{TREC-DL 2019}      & \multicolumn{3}{c}{TREC-DL 2020}   \\
\cmidrule(lr){2-4}\cmidrule(lr){5-7}
 & Latency ($\times$) & nDCG@10 & R@1k & Latency (x) & nDCG@10 & R@1k \\
\midrule
None    & 19.9 & 69.6 & 80.6 & 21.7 & 65.3 & 83.5
   \\
\midrule
\multicolumn{7}{c}{Prune by document size (D)} \\
\midrule
64         & 13.5 (1.5$\times$) & 69.5 & 80.3 & 15.0 (1.5$\times$) & 63.6 & 83.0 \\
32         & 5.7 (3.5$\times$)  & 67.8 & 79.0 & 5.2 (4.2$\times$)  & 62.6 & 81.4 \\
16         & 3.4 (5.9$\times$)  & 64.1 & 77.5 & 3.5 (6.2$\times$)  & 57.3 & 79.0 \\
8          & 1.5 (13.3$\times$)  & 57.6 & 71.5 & 1.5 (14.5$\times$)  & 52.5 & 70.4\\
\midrule
\multicolumn{7}{c}{Prune low posting list scores (T)} \\
\midrule
Prune lower 50\%        & 11.4 (1.7$\times$) & 67.9 & 72.3 & 13.1 (1.7$\times$) & 60.9 & 76.3 \\
Prune lower 75\%        & 10.0 (2.0$\times$) & 62.1 & 60.8 & 7.0 (3.1$\times$) & 51.8 & 61.2 \\
Prune lower 85\%        & 4.9 (4.1$\times$)  & 56.4 & 48.3 & 5.4 (4.0$\times$) & 44.3 & 50.6 \\
\midrule
\multicolumn{7}{c}{Prune low scores (A)} \\
\midrule
Prune $< 0.60$  & 5.2 (3.8$\times$) & 66.2 & 79.9 & 5.4 (4.0$\times$) & 59.9 & 80.8 \\
Prune $< 0.80$  & 2.4 (8.3$\times$) & 63.6 & 77.6 & 2.4 (9.0$\times$) & 55.3 & 78.2 \\
Prune $< 1.00$  & 1.2 (16.6$\times$) & 58.5 & 72.2 & 1.1 (19.7$\times$) & 50.3 & 72.2 \\
Prune $< 1.25$  & 0.5 (39.8$\times$) & 49.8 & 66.5 & 0.5 (43.4$\times$) & 38.7 & 59.7 \\

\bottomrule
\end{tabular}
}
\end{table}

%% file: appendix_tables/TREC/splade.tex
\begin{table}[ht]
\caption{Experiments with eff-V-L SPLADE on the TREC-DL queries (2019 and 2020). The $\times$ value represents the speedup compared to the baseline.}
\label{tab:trec_splade_full}
\adjustbox{max width=\columnwidth}{%
\begin{tabular}{ccccccc}
\toprule
Pruning parameter & \multicolumn{3}{c}{TREC-DL 2019}      & \multicolumn{3}{c}{TREC-DL 2020}   \\
\cmidrule(lr){2-4}\cmidrule(lr){5-7}
 & Latency ($\times$) & nDCG@10 & R@1k & Latency (x) & nDCG@10 & R@1k \\
\midrule
None    & 28.2 (1.0$\times$) & 71.4 & 84.8 & 30.2 (1.0$\times$) & 71.5 & 86.2 \\
\midrule
\multicolumn{7}{c}{Prune by document size (D)} \\
\midrule
64        & 13.2 (2.1$\times$) & 69.3 & 83.5  & 13.7 (2.2$\times$) & 70.4 & 84.8\\
32        & 8.8 (3.2$\times$)  & 67.9 & 79.0  & 8.9 (3.4$\times$) & 67.2 & 81.8\\
16         & 6.0 (4.7$\times$)  & 64.4 & 74.6  & 6.5 (4.6$\times$) & 65.0 & 79.4\\
8         & 3.4 (16.6$\times$) & 58.82 & 67.0  & 3.5 (8.6$\times$) & 59.5 & 70.2\\
\midrule
\multicolumn{7}{c}{Prune low posting list scores (T)} \\
\midrule
Prune lower 50\%        & 15.3 (1.8$\times$) & 70.1 & 83.9 & 15.2 (2.0$\times$) & 70.7 & 85.4 \\
Prune lower 75\%        & 9.2 (3.1$\times$)  & 69.0 & 79.6 & 9.5 (3.2$\times$) & 68.7 & 82.3 \\
Prune lower 85\%        & 5.7 (4.9$\times$)  & 67.5 & 75.1 & 7.5 (4.0$\times$) & 65.8 & 78.9 \\
\midrule
\multicolumn{7}{c}{Prune low scores (A)} \\
\midrule
Prune $< 0.50$       & 14.1 (2.1$\times$)  & 70.1 & 84.0  & 15.1 (2.0$\times$) & 70.8 & 85.3\\
Prune $< 0.75$       & 10.3 (3.2$\times$)  & 68.7 & 82.2  & 10.2 (3.4$\times$) & 69.9 & 84.1\\
Prune $< 1.00$       & 8.4 (4.7$\times$)   & 68.2 & 80.2  & 6.9 (4.7$\times$) & 67.9 & 82.24\\
Prune $< 1.25$       & 5.9 (6.4$\times$)   & 65.4 & 77.9  & 6.3 (8.7$\times$) & 65.7 & 82.0\\

\bottomrule
\end{tabular}
}
\end{table}